\documentclass[prl,reprint,showpacs,twocolumn,superscriptaddress]{revtex4}
\usepackage{graphicx}				
\usepackage{footnote}								
\usepackage{amssymb}

\begin{document}

\author{Timur A. Isaev}
\altaffiliation[on leave from:~]{Petersburg Nuclear Physics Institute NRC KI, 188300, 
Gatchina, Orlova Roscha 1, Russia}
\affiliation{Fachbereich Chemie, Philipps-Universit{\"a}t Marburg, 35032, Marburg, Hans-Meerwein-Str 4, Germany}
\author{ Robert Berger}
\affiliation{Fachbereich Chemie, Philipps-Universit{\"a}t Marburg, 35032, Marburg, Hans-Meerwein-Str 4, Germany}
\date{\today}							
\pacs{31.50.-x, 33.20.Tp, 33.70.Ca, 37.10.Mn}

\title{Polyatomic candidates for cooling of molecules with lasers from simple theoretical concepts}

\begin{abstract}
A rational approach to identify polyatomic molecules that appear to be promising candidates for direct
Doppler cooling with lasers is outlined. First-principle calculations for equilibrium structures and Franck--Condon factors of
selected representatives with different point-group symmetries (including the chiral non-symmetric group C$_1$) have been performed 
and high potential for laser-cooling of these molecules is indicated.
\end{abstract}

\maketitle

\paragraph{Introduction}
Cold molecules are currently heavily studied and serve as particularly interesting targets for spectroscopy 
\cite{Schnell:2009,ChemRev:2012}.
A wealth of applications of cold molecular systems for high-precision spectroscopy including the search for violation of
fundamental symmetries \cite{isaev:2010, Leanhardt:11}, dark matter \cite{Roberts:2014}  and variation of fundamental
constants \cite{Flambaum:2007} as well as controlled cold chemistry \cite{Krems:2008},
quantum computations (see e.g. review \cite{Negretti:2011}) and many others, have been considered recently. 
A special issue of Chemical Reviews \cite{ChemRev:2012} contains comprehensive reviews of some popular techniques 
for obtaining samples of cold and ultra-cold molecules, among which are photoassociation 
from cold atoms, Stark ans Zeeman deceleration of the molecular beams and buffer-gas cooling.
Presently the stimulated Raman adiabatic passage (STIRAP) technique used to transfer Feshbach molecules to the
rovibronic ground state provides lowest temperatures and highest densities of diatomic molecular gases
(see e.g. \cite{Ni:2008} and recent developments in \cite{Wu:2012, Park:2015}). 

Considerable progress has been reported for direct Doppler cooling of diatomic molecules with lasers 
\cite{Shuman:10, Hummon:2013}. The latter method looks especially promising, as it is intimately connected
with well-developed techniques for cooling and trapping of
atoms. Thus, in principle, diverse advances in atomic Doppler cooling can
be transferred to the related schemes for molecules. For example the magneto-optical trapping widely 
used with cold atoms was recently
demonstrated for YO \cite{Hummon:2013} and SrF \cite{Barry:2014}. 

As of yet, however, only diatomic molecules could successfully be Doppler cooled with lasers as
molecules amenable for Doppler cooling must feature Franck--Condon (FC) factors close to unity for
transition between those vibronic states involved in the closed transition cycles (usually the ground and first
excited electronic state). This is connected with the fact that Doppler cooling requires absorption and reemission  
of thousands to millions of photons per atom or molecule involved in cooling. 
The reemission process creates a dissipative force for species counterpropogating the
laser beam, whereas working levels for those species copropogating the laser beam are shifted out of resonance
 because of the Doppler effect. Thus,
closure of the cooling transition is crucial for Doppler cooling.
In general, such closed transitions do not exist in molecules due to non-selective spontaneous decay
from the selected rovibronic level of the excited electronic state to a multitude of rovibrational levels 
of the electronic ground state. 
The probability of such a process is proportional to the FC factor between the corresponding vibrational levels.
As it was first displayed by DiRosa in 2004 \cite{DiRosa:04}, there exist diatomic molecules with the 
sum of a few largest FC factors (for transitions originating from a given vibrational level of the
excited electronic state) being very close to unity. 
Peculiarities of the electronic structure in systems possessing such quasi-diagonal FC matrices
have been identified in Ref. \onlinecite{isaev:2010}. Almost parallel Born--Oppenheimer potential energy curves of the electronic
ground and excited state (which results in diagonality of the FC matrix) are expected in diatomic molecules with one
valence electron over closed shells, when essentially this electron  undergoes transition between (mainly) 
non-bonding orbitals upon electronic excitation. 
As electronic transition between non-bonding orbitals influences the chemical 
bonds (formed by electrons on bonding orbitals) typically only weakly, 
the internuclear potential stays practically unchanged for ground 
and excited electronic states. The influence of the non-bonding electron is even smaller, when due 
to e.g destructive $s$-$p$ hybridization the electron is shifted away from the bonding region (see next 
section).

\begin{figure}[h]
\caption{\label{orbs} (Color online)
Surfaces of constant value (isosurfaces) of the non-bonding one-electron 
wavefunctions (orbitals) of the valence electron for
 CaCH$_3$. The singly occupied molecular orbital (SOMO) for the ground electronic configuration
 is in solid color (light-blue for positive and dark-blue for negative signs of the orbital; the
 iso value is 0.055 $a_0^{-3/2}$), 
 whereas the SOMO for the excited state configuration is in transparent 
 yellow color (light-yellow for positive and dark-yellow for negative signs of the orbital; the
 iso value is 0.040 $a_0^{-3/2}$). It is seen that the electronic centre-of charges are shifted outside of the bonding region 
 and that the ground-state SOMO is mainly a mixture of $s$ and $p$ wavefunctions centered on Ca nucleus 
 ($s$-$p$ -- hybridised orbital), 
 whereas the excited state SOMO is rather $p$-$d$ -- hybridised.}
\includegraphics[width=0.97\linewidth]{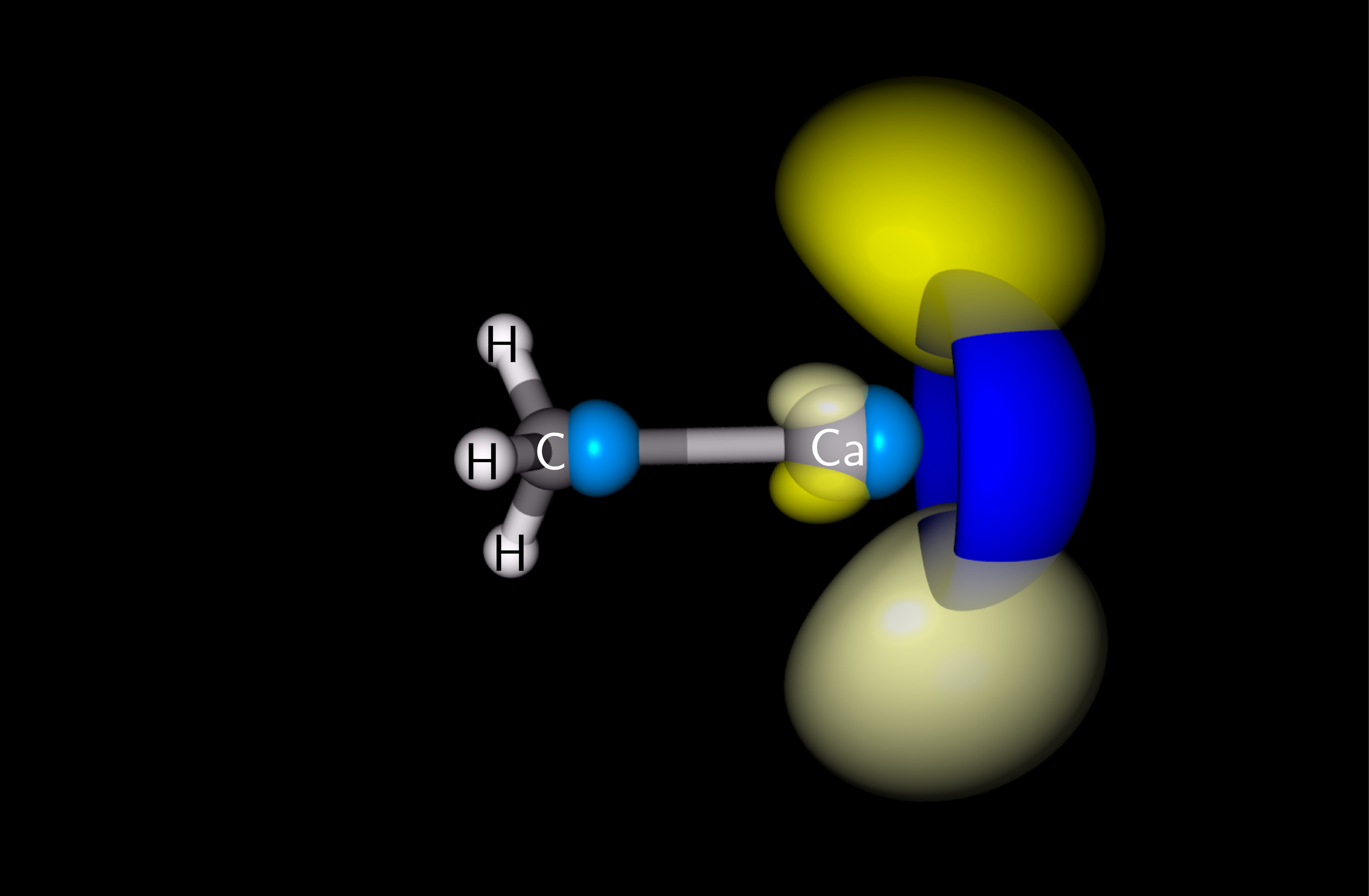}
\end{figure}

\begin{figure*}[h]
\caption{\label{fc-fig} (Color online)
Schematic graphical representation of the vibronic transitions with largest FC factors for CaNC and CaOH molecules.
On coordinate axes the normal coordinates $Q_1$ and $Q_2$ are shown for both ground and excited electronic states. 
$Q_1$ correspond mainly to Ca-X stretching $\Delta R=R-R_\mathrm{e}$, where X=OH, NC, 
$R_\mathrm{e}$ is the equilibrium distance Ca-X from Table I in \cite{polyatomic-supp:2015}. 
Coordinate $Q_2$ is approximately equal to $\theta$-$\theta_\mathrm{e}$, where $\theta$ is the bending angle and 
$\theta_\mathrm{e}$ is its equilibrium value according to Table I in \cite{polyatomic-supp:2015}.
The approximate displacement 
vectors for Ca and CN and OH groups are shown next to corresponding coordinate axes.
All parameters, including the energy $E$, are in arbitrary units (for the sake of clearer representation). 
On both plots the vibrational quantum numbers and overtones are shown next to the corresponding arrows, 
e.g 1$^{0}_{2}$ denotes the vibronic transition
between the ground (0) vibrational state of the excited electronic state and the vibrational state with the first mode (1)
(Ca-X stretching mode) being doubly excited (2) in the ground electronic state.}
\includegraphics[width=0.97\linewidth]{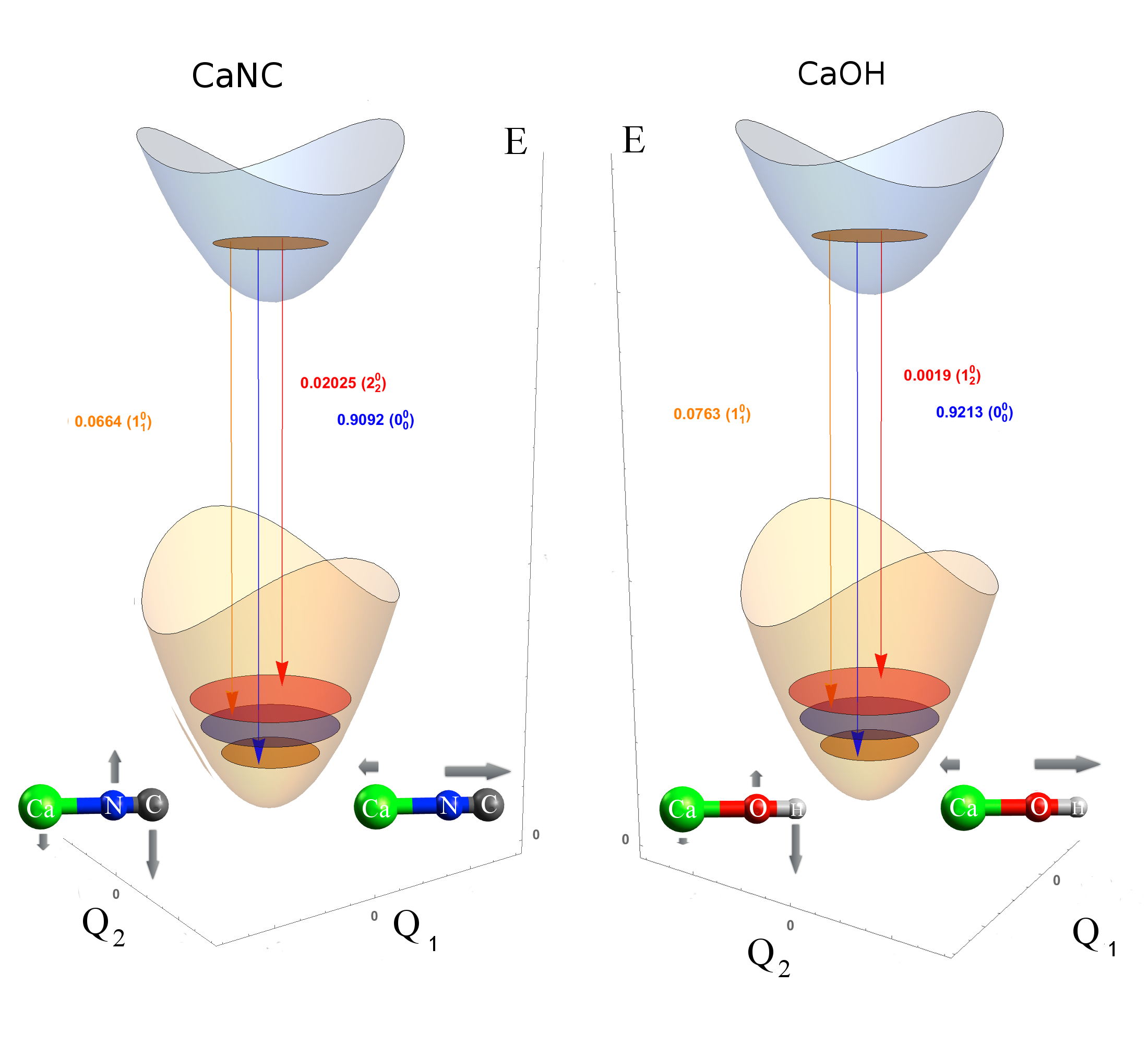}
\end{figure*}

We  pointed out \cite{itamp:2010} that also in polyatomic molecules with a valence electron undergoing transitions between non-bonding orbitals quasi-closed transition loops should be present, which opens up an avenue to cold samples of larger molecules.
In this Letter we provide examples of polyatomic molecules that may be expected to have quasi-diagonal 
FC matrices for vibronic transitions between electronic states and are thus potentially amenable for
direct cooling with lasers. Classes of molecules and explicit examples together with possible routes for
generating proper molecular structures were originally reported on the Conference for Cold and Controlled
Molecules and Ions (Monte Verit{\`a} (Switzerland), 2014).
\paragraph{General consideration}
 A straight-forward way to identify polyatomic molecules with electronic properties similar 
 to those of diatomic molecules is to apply substitution: 
 e.g. in the series of MF open-shell diatomic molecules (class I molecules, see \cite{isaev:2010} and below in this section), 
 where M is a metal atom from group 2 of the Periodic Table (Be-Ra), one can substitute fluorine for a
pseudohalogen such as CN, NC, SCN or functional groups such as OH or CH$_3$.
The pseudohalogens are polyatomic functional groups whose chemical properties 
resembles those of halogens. Particularly the bonding situation in molecules containing 
pseudohalogenic groups is close to their halogen analogs, a fact known
in chemistry for almost a century (see e.g. \cite{Birckenbach:1925}).
From the electronic structure point of view one might expect that the occupation pattern of the leading electronic configurations for 
the ground and energetically lowest excited state of the resulting compounds differs only for the unpaired electron, 
which occupies then different (essentially) non-bonding orbitals, 
similar to the original MF molecule (see also \cite{bernath:1991} for a more detailed discussion on the expected structure for calcium derivatives).  
 For some of the above-mentioned compounds we also benefit from the fact that the respective 
 non-bonding orbitals are mainly centred on the metal atom with their centre-of-charge 
 shifted away from the bonding region (see Fig.\ref{orbs}). 
 Of course, an analogous scheme can be applied to class II  molecules. 
The classification of molecules according to \cite{isaev:2010} is based on electronic structure arguments.
 Class I molecules have non-bonding orbitals which appear due to interference (hybridization)
 of atomic valence one-electron wavefunctions (e.g. $s$-$p$ or $p$-$d$ hybridization as it is shown in 
 Fig.\ref{orbs}). Non-bonding orbitals in Class II molecules appear due to symmetry reasons, like in the
 HI$^+$ molecular ion \cite{Isaev:05a}. We empasize, however, that even in the case of a valence electron undergoing 
 a transition between non-bonding orbitals, quasidiagonality of the FC matrix can only be {\it expected}, but not 
 {\it guaranteed}. 
Counteracting are, for instance, the Renner--Teller or Jahn--Teller effects, which are absent for diatomic
molecules, but can considerably influence the equilibrium structure and vibrational frequencies of polyatomic
molecules.  For example, for the molecule MgOH the equilibrium structure of the electronic ground state 
is linear, whereas the one of the energetically lowest electronically excited state 
is bent \cite{Theodorakopoulos:1999}, so that a quasi-diagonal FC matrix can hardly be expected in this case.
Typically, FC factors are extremely sensitive to relative displacements in equilibrium distances of the electronic
states.  Calculation of FC factors with an accuracy required for a reliable prediction of laser coolability 
is challenging even for diatomic molecules. Nevertheless, even a 10\% accuracy in calculations of FC factors, which can be 
conventionally reached in modern quantum chemical approaches for large (bigger than 0.5) FC factors, 
allows to considerably narrow down 
the initial selection of candidates, whereas simple theoretical concepts primarily provide guidelines
for deducing molecular candidates.
 \paragraph{Linear triatomic molecules} 
 Examples of linear molecules, for which quasidiagonal FC matrices may be expected, are calcium monohydroxide (CaOH) and calcium
monoisocyanide (CaNC). CaOH and CaNC have the advantage of featuring a linear equilibrium structure in the ground and
essentially also in the energetically lowest excited doublet state, thus allowing direct adaptation of the cooling scheme used for
diatomic molecules. Besides, CaNC possesses a large electric dipole moment (about 6 Debye, see Table I
in \cite{polyatomic-supp:2015}),
which renders this molecule especially attractive for the combined scheme of Stark deceleration and subsequent (or
simultaneous) laser cooling. We calculated molecular parameters and FC factors for vibronic
transitions between ground and excited electronic states. The results are summarised in the Table I
in \cite{polyatomic-supp:2015}, whereas FC factors together with vibrational levels are graphically represented on Fig.\ref{fc-fig}.
We considered three largest FC factors as typical cooling experiment of diatomic molecules suggest one main pump transition and two
repumpers as e.g. in \cite{Shuman:10}.
To estimate the Doppler temperature for both CaOH and CaNC we calculated the natural fluorescence lifetime $\tau$ of the excited states
(see Table III in \cite{polyatomic-supp:2015}).
Taking $\tau$ being approximately
equal to 20 ns, we obtain a Doppler temperature $T_\mathrm{D}$ for both CaOH and CaNC 
according to formula $T_\mathrm{D}=\hbar / (2 k_\mathrm{B} \tau)$
where $k_\mathrm{B}$ is the Boltzman constant and $\hbar = h/(2 \pi)$ the reduced Planck constant. $T_\mathrm{D}$ is then about 1 mK, which is 
considerably lower than the typical temperature which can be reached in the framework of e.g. buffer-gas cooling. 
  \paragraph{Computational details:}
 Input files for calculations and additional data are also provided in \cite{polyatomic-supp:2015}.
 The {\sc molpro} program package \cite{molpro2012a} was used for electronic structure calculations.
 Subsequent computation of FC factors in the harmonic approximation were performed with our hotFCHT code
\cite{Berger:1997, Jankowiak:2007, Huh:2012}.   
 Basis sets of quadruple-zeta or triple-zeta quality augmented with polarisation valence basis functions
 (def2-QZVPP and def2-TZVP) 
 were used for all atoms.  All electronic structure calculations started with the
 closed-shell Hartree-Fock calculations of the ground-state configuration of the singly charged 
 molecular cation to obtain an initial guess for the molecular orbitals. 
 Then calculations in the framework of complete active space self-consistent field (CASSCF, \cite{malmqvist:1989} )
were performed for the neutral molecule, followed by configuration interaction (CI, \cite{Werner:1988}) calculations.
The FC factors for transitions between the ground vibrational state of the first electronically excited state and
vibrational states of the electronic ground state are estimated in all cases. A few calculations with
different level of accounting for electronic correlations and basis sets were used to check the stability of the
computed FC factors. Details of the calculations can be found in \cite{polyatomic-supp:2015}. We emphasize that in
all cases the sum of three largest FC factors is always larger than 0.9 for both CaNC and CaOH. This clearly
indicates the potential of these molecules (and, although with exceptions, the general class MOH and MNC, 
where M is an alkaline earth metal Be-Ra) for direct cooling with lasers.
 \paragraph{Non-linear polyatomic molecules}
  Examples of non-linear polyatomic molecules having a valence electron on non-bonding orbitals are CaCH$_3$ 
 and MgCH$_3$. The scheme of arriving at these molecules is analogous to the one in the previous paragraph, with the obvious difference
 in the replacement of the pseudohalogen by methyl. The calculation results are summarized in Table II of \cite{polyatomic-supp:2015}.
 Both molecules have quasi-diagonal 
 FC matrices for transitions between electronic ground and first excited state (see 
 Table II in \cite{polyatomic-supp:2015}). The important feature of these molecules is a quite small Jahn-Teller effect 
 (see the data of the experimental measurements 
 of Jahn-Teller parameters in \cite{Salzberg:1999} for MgCH$_3$), which also favours a quasi-diagonal FC matrix.
 The computational scheme was essentially analogous to the one used for CaOH and CaNC (same basis sets, calculations of the closed-shell cation at first stage followed by MCSCF and CI calculations, see \cite{polyatomic-supp:2015}).
 As detailed and high-precision electronic structure calculations are beyond the scope of the current letter,
 we used a simplified scheme by considering only the totally-symmetric modes (C$_\mathrm{3v}$ molecular symmetry group) 
 in estimation of FC factors rather than solving the full vibronic coupling problem. To check the influence 
 of the non-totally symmetric modes, calculations for MgCH$_3$ were performed without symmetry restrictions and then FC factors were calculated
 in the harmonic approximation. 
Although the calculated molecular parameters are in reasonable agreement with experimental measurements (see Table II), the {\it displacements} of
bond lengths M-C (M=Ca,Mg) predicted from our calculations are larger in absolute value than the experimentally measured ones
(see Tables VII-X in \cite{polyatomic-supp:2015}). Nevertheless, we obtained a sum of FC factors for three vibronic transitions with largest FC factors
 exceeding 0.8 for MgCH$_3$ even when non-totally symmetric modes are accounted for. Thus one can hope for even larger values of FC factors 
 in experiment. It should be emphasized that we primarilly consinder the
 sum of the three larges FC factors, because the value is (according to our observation) more stable to variations in relative displacements than individual FC factors.
 In cases where only totally-symmetric modes are accounted for, the symmetry blocks are not perfectly separated 
 due to some residual symmetry breaking in the excited state structures. Nevertheless, we obtained a sum of the three largest FC factors exceeding 0.9 for both
 MgCH$_3$ and CaCH$_3$. A detailed computational investigation of Jahn--Teller vibronic coupling in these systems shall be left for future studies.
 We note in passing that the quite large electric dipole moment computed for CaCH$_3$ indicates 
 the potential for Stark deceleration. Doppler temperatures for MgCH$_3$ and CaCH$_3$ can be estimated analogously to linear molecules,
 using calculated lifetimes of the excited states from Table III. Our order-of-magnitude estimate of $T_\mathrm{D}$ 
 is the same as for linear molecules, about 1 mK.   
 \paragraph{Chiral molecules}
Cooling and trapping of chiral molecules would mean a crucial improvement for spectroscopic measurements
of a number of highly interesting effects e.g. parity violating energy differences between enantiomers due to the fundamental weak interaction 
\cite{berger:2004a, quack:1989}. Using the scheme from the above paragraphs it is also possible to identify  chiral molecular
structures having a quasidiagonal FC matrix. One way is isotopic substitution to a known molecule
\cite{berger:2004a, berger:2005b}. The chiral methyl group CHDT is well established in chemistry (see 
review \cite{Floss:1993} for general aspects as well as ref.~\cite{berger:2003a} on parity violation in CHDTOH).
We calculated FC factors for the molecule MgCHDT, using the same method for electronic
structure calculation as in the case of MgCH$_3$, without accounting for molecular symmetry. The resulting FC factors we obtained are
as follows: (0.64, 0.18, 0.04) with $\sum$ equal to 0.86.  Thus FC factors are predicted to be only slightly
smaller than for MgCH$_3$. This modest change is
excellent news for attempts to obtain (ultra)cold chiral molecules for the search of violation of fundamental
symmetries.
\paragraph{Conclusions}
We have outlined a rational approach to identify polyatomic molecules that appear to be promising candidates for direct
Doppler cooling with lasers. Explicit numerical calculations for structures and Franck--Condon factors of
selected representatives indicate high potential for laser-cooling of such molecules to even open up the third spatial
dimension for ultra-cold molecules generated by direct Doppler cooling with lasers. 
\begin{acknowledgments}
We are very grateful to organisers (Christiane Koch and Andreas Osterwalder) and participants of the Conference for Cold and Controlled Molecules and Ions
(Monte Verit{\`a} (Switzerland), 2014) for fruitful discussions and critical remarks. In particular we thank Ronnie Kosloff, who attracted our attention to 
CaNC. Hauke Westemeier is acknowledged for creating the figure with the molecular orbitals. We thank the referees 
of our MS published in Phys. Rev. Letters for fruitful and detailed 
critical remarks, especially concerning the representation of our results.
\end{acknowledgments}
%

\end{document}